\documentclass[fleqn,usenatbib,useAMS]{mnras}

\usepackage{newtxtext,newtxmath}

\usepackage[T1]{fontenc}

\DeclareRobustCommand{\VAN}[3]{#2}
\let\VANthebibliography\thebibliography
\def\thebibliography{\DeclareRobustCommand{\VAN}[3]{##3}\VANthebibliography}

\usepackage{graphicx}	
\usepackage{amsmath}	
\usepackage{ulem}
\usepackage{xcolor}
\usepackage{soul}

\title[High-resolution ALMA observations of V4046\,Sgr]{High-resolution ALMA observations of V4046\,Sgr: a circumbinary disc with a thin ring}

\author[R. Martinez--Brunner et al.]{Rafael Martinez--Brunner,$^{1}$
\thanks{E-mail: rmartinezbrunner@gmail.com}
Simon Casassus,$^{1}$
Sebasti\'an P\'erez,$^{2,3}$
Antonio Hales,$^{4,5}$
Philipp Weber, $^{1,2,3}$ \newauthor
Miguel Carcamo,$^{3,6,7}$
Carla Arce-Tord,$^{1}$
Lucas Cieza,$^{8}$
Antonio Garufi,$^{9}$
Sebasti\'an Marino,$^{10}$
Alice Zurlo$^{8,11,12}$
\\
$^{1}$Departamento de Astronom\'ia, Universidad de Chile, Casilla 36-D, Santiago, Chile\\
$^{2}$Departamento de F\'isica, Universidad de Santiago de Chile, Av. Victor Jara 3659, Santiago\\
$^{3}$Center for Interdisciplinary Research in Astrophysics and Space Exploration (CIRAS), Universidad de Santiago de Chile, Estaci\'on Central, Chile \\
$^{4}$Joint ALMA Observatory, Avenida Alonso de C\'ordova 3107, Vitacura 7630355, Santiago, Chile \\
$^{5}$National Radio Astronomy Observatory, 520 Edgemont Road, Charlottesville, VA 22903-2475, United States of America \\
$^{6}$Departamento de Ingenier\'ia Inform\'atica, Universidad de Santiago de Chile, Av. Ecuador, 3659, Santiago, Chile\\
$^{7}$Jodrell Bank Centre for Astrophysics, Alan Turing Building, Department of Physics and Astronomy, Alan Turing Building,\\ University of Manchester, Oxford Road, Manchester, M13 9PL, UK \\
$^{8}$N\'ucleo de Astronom\'ia, Facultad de Ingenier\'ia y Ciencias, Universidad Diego Portales, Av. Ejercito 441, Santiago, Chile\\
$^{9}$INAF, Osservatorio Astrofisico di Arcetri, Largo Enrico Fermi 5, I-50125, Firenze, Italy\\
$^{10}$Institute of Astronomy, University of Cambridge, Madingley Road, Cambridge CB3 0HA, UK\\
$^{11}$Escuela de Ingenier\'ia Industrial, Facultad de Ingenier\'ia y Ciencias, Universidad Diego Portales, Av. Ejercito 441, Santiago, Chile \\
$^{12}$Aix Marseille Universit\'e, CNRS, LAM - Laboratoire d'Astrophysique de Marseille, UMR 7326, 13388, Marseille, France  \\
}

\date{Accepted XXX. Received YYY; in original form ZZZ}

\pubyear{2021}

\begin{document}
\label{firstpage}
\pagerange{\pageref{firstpage}--\pageref{lastpage}}
\maketitle

\begin{abstract}
    The nearby V4046\,Sgr spectroscopic binary hosts a gas-rich disc known for its wide cavity and dusty ring. We present high resolution ($\sim$20\,mas or 1.4\,au) ALMA observations of the 1.3\,mm continuum of V4046\,Sgr which, combined with SPHERE--IRDIS polarised images and a well-sampled spectral energy distribution (SED), allow us to propose a physical model using radiative transfer (RT) predictions. The ALMA data reveal a thin ring at a radius of 13.15$\pm$0.42\,au (Ring13), with a radial width of 2.46$\pm$0.56\,au. Ring13 is surrounded by a $\sim$10\,au-wide gap, and it is flanked by a mm-bright outer ring (Ring24) with a sharp inner edge at 24\,au. Between 25 and $\sim$35\,au the brightness of Ring24 is relatively flat and then breaks into a steep tail that reaches out to $\sim$60\,au. In addition, central emission is detected close to the star which we interpret as a tight circumbinary ring made of dust grains with a lower size limit of 0.8\,mm at 1.1\,au. In order to reproduce the SED, the model also requires an inner ring at $\sim$5\,au (Ring5) composed mainly of small dust grains, hiding under the IRDIS coronagraph, and surrounding the inner circumbinary disc. The surprisingly thin Ring13 is nonetheless roughly 10 times wider than its expected vertical extent. The strong near-far disc asymmetry at 1.65\,$\micron$ points at a very forward-scattering phase function, and requires grain radii of no less than 0.4\,$\micron$.
\end{abstract}




\section{Introduction} \label{sec:Introduction}

Recent observations of young circumstellar discs have transformed the current knowledge on planet formation. Among the main recent findings of resolved observations of discs is the discovery of substructures in the form of gaps, rings, cavities, and spirals, to name a few \citep[see,][and references there in]{Andrews_2020_aug}. However, the focus of resolved imaging with the Atacama Large Millimeter/submillimeter Array (ALMA) or with the current generation of high-contrast cameras has mainly been towards the brighter sources \citep[e.g.,][]{2017A&A...603A..21G}. It is to address this bias that the ``Discs Around T Tauri Stars with SPHERE'' (DARTTS-S) programme collected differential polarization imaging (DPI) data with the Spectro-Polarimeter High-contrast Exoplanet REsearch \citep[SPHERE][]{2019A&A...631A.155B} for a total of 29 solar-type stars \citep[][]{Avenhaus_2018,Garufi2020}. The sample is not biased towards exceptionally bright and large discs. The DARTTS observations revealed diverse structures and morphologies in the scattering surface of these discs. This article on V4046 Sagittarii (Sgr) is the first instalment of a companion programme, the DARTTS survey with ALMA (DARTTS-A), which will present millimetre observations of nine protoplanetary discs previously imaged in polarised scattered light in DARTTS-S.

V4046\,Sgr is a double-lined spectroscopic binary of K-type stars (K5 and K7) with very similar masses of 0.90$\pm$0.05\,M$_{\sun}$ and 0.85$\pm$0.04\,M$_{\sun}$ \citep{Rosenfeld_2012}, on a close ($a \approx 0.041$\,au), circular ($e\leq0.01$) orbit, with an orbital period of 2.42\,days \citep{2000IAUS..200P..28Q}. It is a member of the $\beta$ Pictoris moving group \citep{Zuckerman_2004}, with an estimated age of 18.5$^{+2.0}_{-2.4}$\,Myr \citep{2020A&A...642A.179M}, and its distance is 71.48$\pm$0.11\,pc \citep{gaiacollaboration2021gaia}. V4046\,Sgr hosts a massive ($\sim$0.1\,M$_{\sun}$) circumbinary disc extending to $\sim$300\,au \citep{Rosenfeld_2013, Rodriguez_2010}, rich in diverse molecular lines \citep{Kastner_2018}.

Previous analysis of radio observations taken with the Sub-Millimeter Array (SMA) \citep{Rosenfeld_2013} and later with ALMA \citep{Guzman_2017,Huang_2017,Bergner_2018,Kastner_2018}, showed that the millimetre dust of the V4046 Sgr circumbinary disc features a large inner hole of $\sim$30\,au and a narrow ring centred around 37\,au that account for most of the dust mass. More recently, \citet{Francis_2020}, with higher definition ALMA images, detected a structured outer disc with an inner disc in the millimetre continuum. On the other hand, by using SPHERE polarized light observations in the J- and H-band, \citet{Avenhaus_2018} confirmed the double-ring morphology previously reported by \citet{Rapson_2015} using GPI data. In these polarized light observations, the surface brightness presents a $\sim$10\,au-wide inner cavity, a narrow ring at $\sim$14\,au, and an outer ring centred at $\sim$27\,au. Later, \citet{Ru_z_Rodr_guez_2019} elaborated on the characterisation of the observed rings in the SPHERE images, more on that in Section~\ref{subsec:SPHERE}.

A parametric model that simultaneously fits the NIR scattered light and millimetre continuum emission links the observations to the structures in the underlying dust and gas density distributions, and sheds light on the complex processes that shaped them. In this paper, we present new ALMA observations of the continuum emission at 1.3\,mm of V4046\,Sgr with unprecedented angular resolution in this source, which reveal an inner ring in the disc. We also reproduce these observations with a 3D parametric model that fits all available data, including the polarized scattered light SPHERE image, the spectral energy distribution (SED), as well as the new high definition 1.3\,mm continuum map. Section~\ref{sec:Observations} describes the available observations that we aim to model. Section~\ref{sec:model} describes the structural parameters of our model. A description of our parameter space exploration can be found in Appendix~\ref{sec:Appendix}. The results of our modelling are presented in Section~\ref{sec:results}, including a discussion of the main findings. Finally, in Section~\ref{sec:Conclusions} we present our conclusions.

\section{Observations} \label{sec:Observations}
\subsection{ALMA}  \label{subsec:ALMA}

New ALMA observations of V4046\,Sgr were obtained in 2017 as part of the Cycle 5 program 2017.1.01167.S (PI: S. Perez). The observations acquired simultaneously the 1.3\,mm continuum and the J = 2$-$1 line of $^{12}$CO (i.e. with a band 6 211-275\,GHz correlator setup). A log of the observations is shown in Table~\ref{tab:Summary_ALMA}. The ALMA array was in its C43-8/9 configuration, with baselines ranging from 92 to 13894\,m which translate into a synthesized beam of $0\farcs062 \,\times \, 0\farcs055$, in natural weights. Here we focus on the continuum observations only. \citet{Francis_2020} included a subset of this dataset, corresponding to the C43-8 configuration, as part of a large sample of transition discs, with a focus on the statistical properties of the inner discs.

\begin{table*}
 \caption{Summary of the new ALMA observations presented in this work. The table shows the total number of antennas, total time on  source (ToS), target average elevation, mean precipitable water vapor column (PWV) in the atmosphere, minimum and maximum baseline lengths and maximum recoverable scale (MRS).}
 \label{tab:Summary_ALMA}
 \begin{tabular}{lccccccccc}
  \hline
  Configuration & Execution Block & N Ant. & Date & ToS & Avg. Elev. & Mean PWV & Baseline & MRS \\
   & (uid://A002/) & & & (sec) & (deg) & (mm) & (m) & (")\\
  \hline
  C43-8/9  & Xc6b674/X44f & 48 & 2017-11-10 & 507 &  50.9 &  1.0  & 113.0 - 13894.4 & 0.571 \\
  C43-8  & Xc72427/X34f2 & 43 & 2017-11-23 & 507 & 45.2 & 0.4  &91.6-8547.6 & 0.8\\
  \hline
 \end{tabular}
\end{table*}

Image synthesis of the ALMA continuum was performed with the {\tt uvmem} package \citep{2006ApJ...639..951C, 2018A&C....22...16C}, which fits a non-parametric model image $I^m_j$ to the data by comparing the observed and model visibilities, $V^\circ_k$ and $V^m_k$, using a least-squares figure of merit $L$:
\begin{equation}
  L = \sum_{k=1}^N \omega_k  |V^\circ_k - V^m_k|^2 + \lambda S,
\end{equation}
where $\omega_k$ are the visibility weights and $\lambda$ is a dimensionless parameter that controls the relative importance of the regularization term $S$.

The chosen regularization term $S$ for this case was the standard image entropy, or
\begin{equation}
S = \sum_{j=1}^M \frac{I_j^m}{M} \ln\left(\frac{I_j^m}{M}\right),
\end{equation}
where $M$ is the default pixel intensity value and is set to $10^{-3}$ times the theoretical noise of the dirty map (as inferred from the visibility weights $\omega_k$). Here we set $\lambda = 0.01$. Similar applications of {\tt uvmem} in the context of protoplanetary discs can be found, for example, in \citet{Casassus2013Natur, 2018MNRAS.477.5104C, Casassus2019MNRAS.483.3278C, Perez2019AJ....158...15P} and in \citet{2020ApJ...889L..24P} using long baseline data. An advantage of {\tt uvmem} compared to more traditional imaging strategies, such as provided by the {\tt tclean} task in CASA, is that the effective angular resolution of the model image is $\sim$3 times finer than the natural weights clean beam \citep[][]{2018A&C....22...16C}, giving us a new approximate {\tt uvmem} resolution of $\sim$0$\farcs021 \,\times \,\sim$0$\farcs018$. This angular resolution is comparable to uniform or super-uniform weights in {\tt tclean}, but it preserves the natural-weights point-source sensitivity. The RMS noise of the {\tt uvmem} image is 26\,$\micron$Jy per resolution beam, at a frequency of 237\,GHz.

The resulting {\tt uvmem} image shown in the top right panel of Fig.~\ref{fig:images_vs_simulated} reveals new substructure of the disc, in the form of two rings of large dust grains with a broad gap between them, i.e. Ring13 at around $\sim$13\,au and Ring24 starting at around $\sim$24\,au (see below for a precise estimation). The wide and bright Ring24 reaches its peak intensity at $\sim$30\,au, continues relatively flat and then breaks at $\sim$35\,au into a steeper tail. While this is the first observation of Ring13, \citet{Ru_z_Rodr_guez_2019} anticipated its existence as their ALMA continuum image showed a distinct excess between $\sim$10 and 17\,au.

Ring13 is surprisingly narrow and seems to be off-centred relative to the Gaia stellar position, at the origin of coordinates in Fig.~\ref{fig:images_vs_simulated}. We determined the centre of the ring and its orientation using the {\tt MPolarMaps} package described in \citet{10.1093/mnras/stab2359}, which minimises the dispersion of the intensity radial profiles between two given radii and returns the optimal values for the disc position angle (PA), inclination, and disc's centre. In an initial optimization we focus on the orientation of Ring13, and choose a radial range from 6 to 18\,au, which covers Ring13 but excludes Ring24. The resulting disc orientation is set at a PA of 257.31$\pm$0.03\,deg (east of north), with an inclination of 147.04$\pm$0.02\,deg, and the optimal ring centre is at $\Delta \alpha = -4\pm0.02$\,mas $\Delta \delta = 13\pm0.05$\,mas relative to the Gaia position of the stars. The offset of the centre of the cavity and the nominal stellar positions are coincident within the pointing accuracy of ALMA, which  is $\sim$5\,mas for the signal to noise ratio of our image (calculated using the ALMA Technical Handbook).

\begin{figure}
    \includegraphics[width=\columnwidth]{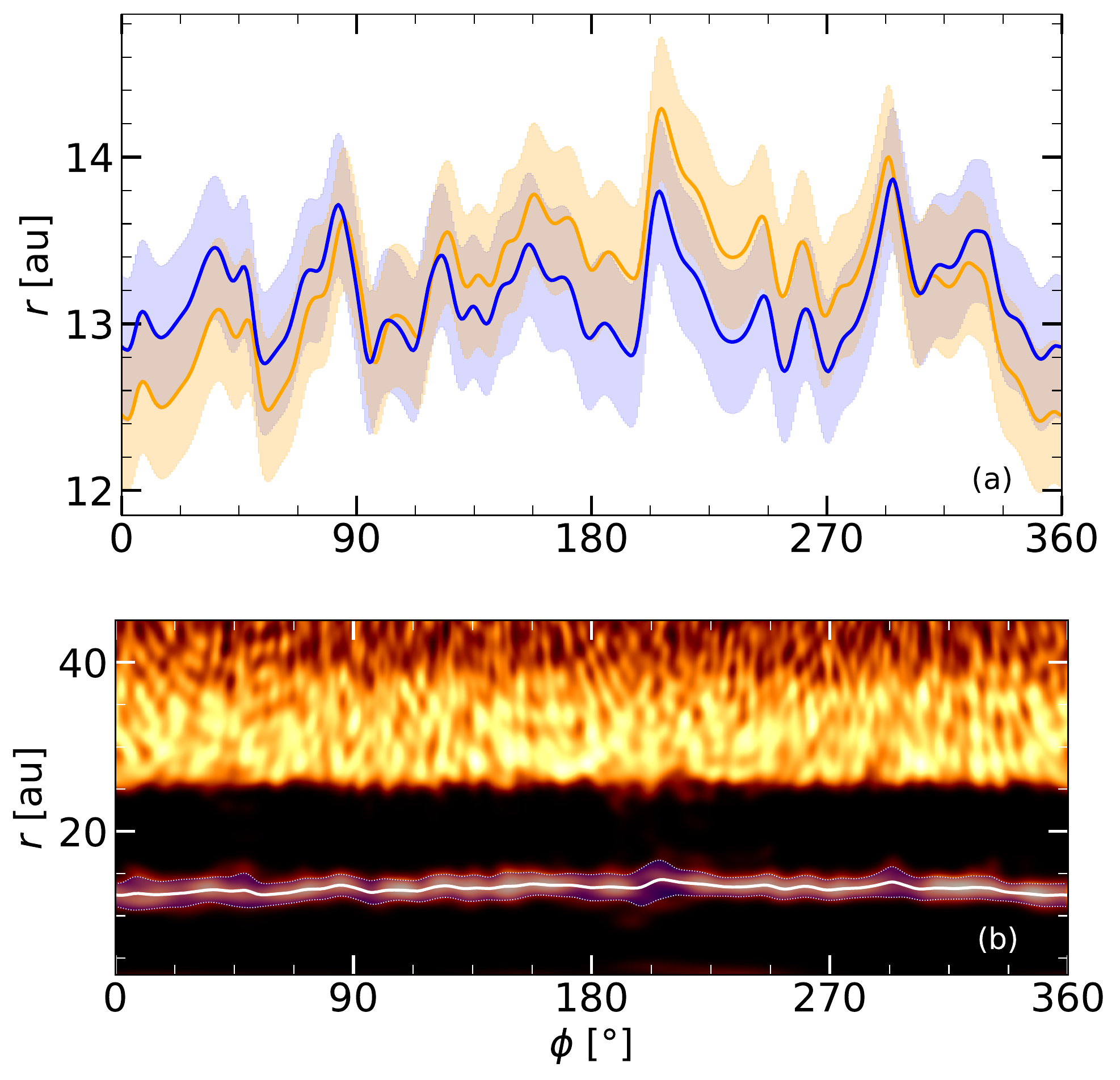}
    \caption{{\bf (a)} Centroid of Ring13, for two disc orientations: the orange line corresponds to the same trace as in (b), while the blue line is obtained for the inner ring orientation. {\bf (b)} Polar decomposition of the 1.3\,mm continuum image, using the orientation of Ring24. We trace Ring13 using the centroids (solid line) and FWHM of radial Gaussian fits (blue region between the dotted lines).}
    \label{fig:polarring}
\end{figure}

In a second optimization of the disc orientation, but this time aiming for Ring24 with a radial domain from 20\,au to 70\,au (fully including Ring24 and excluding Ring13), we obtained a PA of 256.86$\pm$0.02\,deg, with an inclination of 146.08$\pm$0.01\,deg and a centre at $\Delta \alpha = 2\pm0.02$\,mas $\Delta \delta = 12\pm0.02$\,mas relative to the stars. We see that both Ring13 and Ring24 share a very similar orientation and centre, given the errors and the pointing accuracy. However, see that there are some hints for a somewhat different orientation in the azimuthal profiles for the ring radii, which may nonetheless be accounted for by the joint effect of all these small differences in disc orientation. This is summarised in Fig.~\ref{fig:polarring}a, which characterises the radial position and width of Ring13 as a function of azimuth by using radial Gaussian fits at each azimuth. The radial centroids for both disc orientation overlap within the errors, but there is a systematic trend for the difference between the two. It may be that this small difference reflects a finite  intrinsic eccentricity of one or both rings. Deeper imaging is required to progress on this question.

On average, we obtained a radial FWHM for Ring13 of 2.83$\pm$0.50\,au, and a stellocentric radius of 13.15$\pm$0.42\,au (see Fig.~\ref{fig:polarring}b). As the {\tt uvmem} model image has an approximate {\tt uvmem} beam of $\sim$0$\farcs021 \,\times \,\sim$0$\farcs018$ (or $\sim$1.4\,au at 71.48\,pc), we see that Ring13 is resolved. After subtraction of the {\tt uvmem} beam, the ring width is $\sim$2.46$\pm$0.56\,au.

Interestingly, the ALMA image also detects 1.3\,mm continuum emission near the stellar positions (see the inset in Fig.~\ref{fig:images_vs_simulated}). Since this central emission is larger than the angular resolution, it is probably stemming from thermal emission from large dust grains rather than directly from the stars. The peak intensity of this dust structure is at an estimated distance of only $0\farcs012\pm0\farcs002$, or $\sim$0.85$\pm$0.14\,au from the binary system, and, estimating its form as a Gaussian ellipse, it has a mean FWHM of $\sim$1.43\,au.


\subsection{VLT/SPHERE--IRDIS}  \label{subsec:SPHERE}

V4046\,Sgr was observed in DPI mode with SPHERE--IRDIS on March 13, 2016 \citep[see][for details]{Avenhaus_2018}. Here we use a new reduction of the $H$ band data produced with the IRDAP pipeline \citep{2020A&A...633A..64V}, which can separate stellar and instrumental polarization. The polarized signal is consistent with the previous image in \citet{Avenhaus_2018}. The degree of linear polarization of the central and unresolved signal in V4046\,Sgr is only 0.13\%, with a systematic uncertainty of 0.05\% due to time-varying atmospheric conditions during the exposures. The angle of polarization is aligned with the disc major axis, as expected given that the target has an extinction of Av=0.0 \citep{2016ApJ...828...69M} and the entire polarization is dominated by circumstellar rather than inter-stellar material.

\begin{figure*}
  \includegraphics[width=\textwidth]{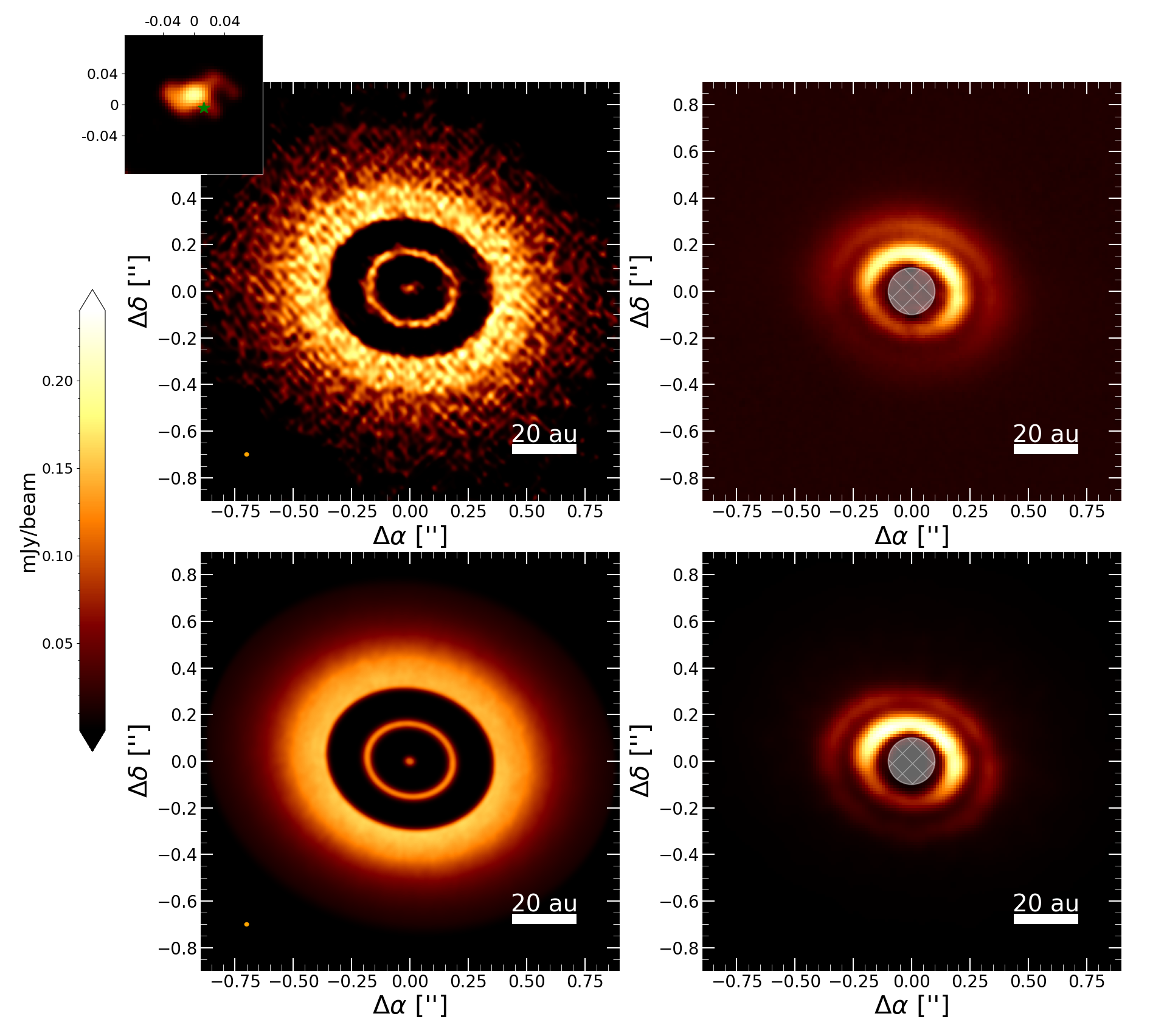}
  \caption{Comparison of observations (top) and simulated images (bottom) at 1.3\,mm continuum (left) and 1.65\,$\micron$ (right) of the circumbinary disc orbiting V4046\,Sgr. \textit{Top left panel:} 1.3\,mm continuum {\tt uvmem} model image. The small orange ellipse shows an estimated {\tt uvmem} beam size ($\sim$0$\farcs021 \, \times \, \sim$0$\farcs018$). The inset zooms into the central emission, and the green star marks the centre of the inner ring. \textit{Top right panel:} SPHERE--IRDIS \textit{H}-band image with a white filled circle that represents the N\_ALC\_YJH\_S coronagraph with a radius of $\sim$0$\farcs12$, or $\sim$8.6\,au at 71.48\,pc.  \textit{Bottom left panel:} synthetic image at 1.3\,mm convolved with the {\tt uvmem} beam. \textit{Bottom right panel:} synthetic image at 1.65\,$\micron$. For all the images in the figure the colour scale is linear, and the color bar on the left applies only to the images on the left-hand panel.}
  \label{fig:images_vs_simulated}
\end{figure*}

The scattered light image in the top left panel of Fig.~\ref{fig:images_vs_simulated} also shows a double ring structure in the micron-sized dust distribution. The observed morphology presents an inner cavity of $\sim$10\,au in radius and two rings located at 14.10$\pm$0.01\,au, coincident with Ring13, and 24.62$\pm$0.08\,au, coincident with the inner wall of Ring24, with a small gap between them at $\sim$20\,au \citep{Ru_z_Rodr_guez_2019}. Two other important features that are present in the image are: the near-far brightness asymmetry, and the shadows projected on the disc by the close binary system as they eclipse each other, discovered by \citet{dOrazi}.

The binary phase reported by \citet{dOrazi} in the scattered light observation is at a PA of 265\,deg, east of north. Using this measurement, the binary phase was calculated at the time of the ALMA observation at a PA of $\sim$80\,deg.

\subsection{Spectral energy distribution} \label{subsec:SED}

\begin{figure}
	\centering
	\includegraphics[width=\columnwidth]{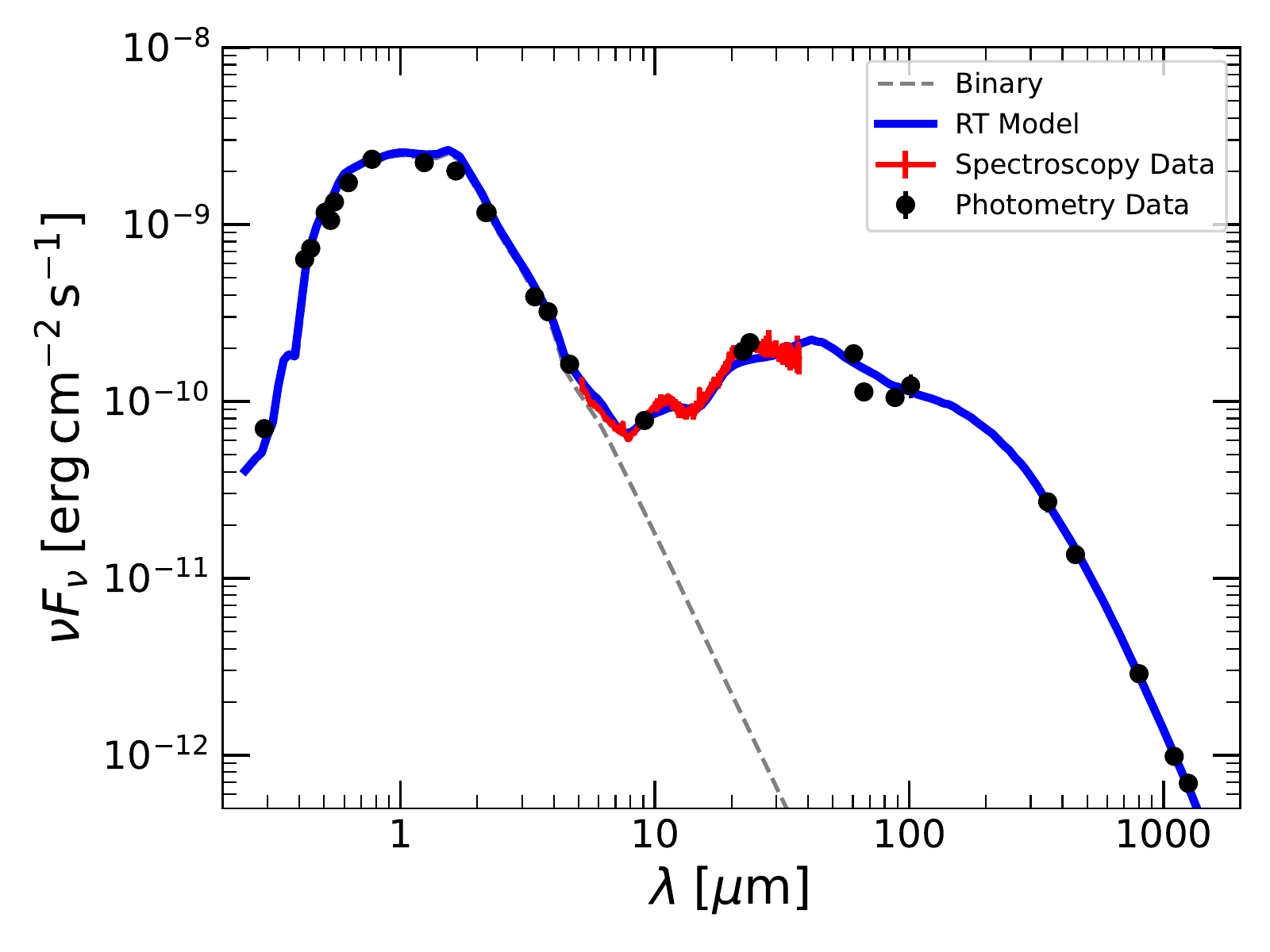}
    \caption{The observed SED of V4046\,Sgr (black points and solid red curve) compared with the model SED resulting from the RT model (solid blue line). The black points represent the measured photometry and the red line shows an archival \textit{Spitzer} IRS spectrum. The dashed silver curve shows the emission of the stellar photosphere model.}
    \label{fig:SED}
\end{figure}

The observed SED was collected from data in the literature \citep{1988iras....7.....H, 1990A&A...234..230H, Jensen_97, 2000A&A...355L..27H, 2001KFNT...17..409K, 2003yCat.2246....0C, 2007PASJ...59S.369M, 2008PASP..120.1128O, 2010A&A...514A...1I, 2012yCat.2311....0C}, available online in \textsc{vizier}. We also used archival \textit{Spitzer} IRS spectroscopic data available in the CASSIS database \citep{Lebouteiller_2015}. The data is displayed in Fig.~\ref{fig:SED} along with the resulting SED of the radiative transfer model presented in the next section (more on the resulting SED in Sec.~\ref{sec:results}). The SED exhibits the dip near 10\,$\micron$ characteristic of transition discs, as described by \citet{Rosenfeld_2013}. \citet{Jensen_97} concluded that these data matched that of a extended circumbinary disc truncated at $\sim$0.2\,au, as the interior would also be expected to be cleared by dynamical effects of the central binary \citep{Art_Lu}. 

\section{Parametric radiative transfer model} \label{sec:model}

The multi-frequency data can be interpreted in terms of a physical structure using radiative transfer predictions, for which we used the \textsc{radmc3d} package \citep[version 2.0,][]{Dullemond_2012}. The general framework of the parametric model that we developed is similar to that in \citet{2018MNRAS.477.5104C} for DoAr\,44, and the initial model values were inspired from those in \citet{Rosenfeld_2013}, \citet{Ru_z_Rodr_guez_2019} and \citet{2019ApJ...882..160Q}. A high-resolution radiative transfer model that reproduces multi-frequency imaging and the SED, is a solution to a highly degenerate problem, so a full parameter exploration requires a level of computation and time that exceeds our capabilities and the scope of this paper. Consequently, our approach was to find through trial and error a set of values for the parameters that closely fit the available data, and then, improve this fit by implementing one dimensional least squared optimizations for some key parameters (more on this in Sec.~\ref{sec:results}). The final structure of the parametric model is summarised in Fig.~\ref{fig:profiles}, where we show on the top panel the surface density profiles for gas and dust grains populations, and in the lower panel the respective aspect ratio profile.

\subsection{General setup}

The stars were modelled using two Kurucz photosphere models \citep{1979ApJS...40....1K, 1997A&A...318..841C}, with T$_{\mathrm{eff},1} =$ 4350\,K, R$_{*,1} =$ 1.064\,R$_{\sun}$, M$_{*,1} =$ 0.90\,M$_{\sun}$ and T$_{\mathrm{eff},2} =$ 4060\,K, R$_{*,2} =$ 1.033\,R$_{\sun}$, M$_{*,2} =$ 0.85\,M$_{\sun}$, respectively and with an accretion rate of $\mathrm{log}(\,\dot{\mathrm{M}}/(\mathrm{M}_{\sun}\,\mathrm{yr^{-1}})) = -$9.3 for both cases to include excess UV due to stellar accretion \citep{10.1111/j.1365-2966.2011.19366.x}. The stars were placed at a mutual separation of 0.041\,au, so that their centre of mass coincides with the origin. Their PA was set to 250\,deg, so that the secondary lies to the NW from the primary, thus casting the same shadow as observed by \citet{dOrazi}.

Reproducing the radial and vertical structure of the V4046\,Sgr disc turned out to be challenging. We built the model in terms of the gas distribution, and with two main dust populations: large grains with radii from 0.3\,$\micron$ to 10\,mm that are vertically settled and dominate the total dust mass, and a population of smaller grains with radii ranging from 0.3 to 1.5\,$\micron$ that are uniformly mixed with the gas and reach higher regions above the mid-plane.

We take a three dimensional model in a cylindrical reference frame with coordinates ($r$, $\theta$, $z$). The inner radius of the model grid was set to 0.1\,au, and the outer radius to 100\,au, which is large enough for the dust disc to be undetectable. We set the values of the inclination and disc position angle to the same as obtained from the ALMA observation in Section~\ref{sec:Observations}, such that the model has an inclination of i = 147.04\,deg and a P.A. = 257.31\,deg. The model was set at a distance of d = 71.48\,pc \citep{gaiacollaboration2021gaia}.

\subsection{Radial structure: gas \& small dust grains}

Given the cylindrical coordinates ($r$, $\theta$, $z$), the gas density ($\rho_{\mathrm{gas}}$) distribution follows
\begin{equation}
  \rho_{\mathrm{gas}}(r,z) =\frac{\Sigma_{\mathrm{gas}}(r)}{\sqrt{2\pi} \, H(r)} \mathrm{exp}\left[-\frac{1}{2} \left(\frac{z}{H(r)}\right)^2\right],
\end{equation}
where $H(r)$ is the scale height profile and $\Sigma_{\mathrm{gas}}(r)$ is the gas surface density profile.

Although both ALMA and SPHERE--IRDIS images display two-ringed morphologies, we propose a three-ringed structure plus an inner disc to reproduce the observations. This bold decision is due the major fit improvement in the SED and the polarize image (more on this in Sec.~\ref{sec:results} and Appendix~\ref{sec:A2}). We separate the gas disc into four individual regions: an inner disc with a power-law profile and three rings named Ring5, Ring13, and Ring24, as they are located at 5, 13, and 24\,au respectively. The combined gas surface density profile is then given by:
\begin{equation}
  \Sigma_{\mathrm{gas}}(r) = \Sigma_{\mathrm{inner\,disc}}(r) + \Sigma_{\mathrm{R5}}(r) + \Sigma_{\mathrm{R13}}(r) + \Sigma_{\mathrm{R24}}(r).
\end{equation}

First, the inner disc model follows a power-law function defined by
\begin{equation}
  \Sigma_{\mathrm{inner\,disc}}(r) =\Sigma_\mathrm{c} \left(\frac{r}{R_\mathrm{c}}\right)^{-\gamma}  \, \mathrm{exp}\left[-\left(\frac{r}{R_\mathrm{c}}\right)^{2-\gamma}\right],
\end{equation}
where $R_c$ is a characteristic radius and $\gamma$ is the surface density power-law index. We used $R_c$ = 16\,au, $\Sigma_\mathrm{c} =1.3\times10^{-4}$\,$\mathrm{g\,cm^{-3}}$ and a fixed $\gamma$ = 1 as it is a typical value for discs \citep{Andrews_2009,Andrews_2010}. The gas in our model extends from $R_{\mathrm{in}}$ = 0.2\,au outwards, consistent with the inner edge radius inferred from the SED data (see Appendix~\ref{sec:A2}).

Secondly, due to the thin nature of Ring5 and Ring13, we chose to use Gaussian profiles to parametrize them,
\begin{equation}
  \Sigma_{\mathrm{ring}}(r) = \frac{\Sigma_\circ}{\sqrt{2 \pi} \sigma}
  \, \mathrm{exp}\left[-\frac{1}{2}\left(\frac{r-\mu}{\sigma}\right)^{2}\right],
\end{equation}
where we define constants that correspond to the centroid radii \{$\mu_{\mathrm{R5}}=5.2$\,au, $\mu_{\mathrm{R13}}=14.9$\,au\}, ring widths \{$\sigma_{\mathrm{R5}}=0.25$\,au, $\sigma_{\mathrm{R13}}=2.26$\,au\}, and normalizations \{$\Sigma_{\circ,\mathrm{R5}}=3.3$\,$\mathrm{g\,cm^{-3}}$, $\Sigma_{\circ,\mathrm{R13}}=6.0 \times 10^{-1}$\,$\mathrm{g\,cm^{-3}}$\} for both components separately.

Thirdly, for Ring24 we used the same power-law as for the inner disc but scaled by an empirically obtained factor, $\delta_{\mathrm{sd}}(r)$ and by $\epsilon(r)$, a parameter that allows us model a smoother inner edge of the outer ring:
\begin{equation}
  \Sigma_{\mathrm{R24}}(r) = \Sigma_{\mathrm{inner\,disc}}(r)\, \delta_{\mathrm{sd}}(r)\,\epsilon(r),
\end{equation}
with $\delta_{\mathrm{sd}}(r) = 1.0\times 10^5$ for $r > 18$\,au and zero for lower radii, and
\begin{equation}
    \epsilon(r) = 
    \begin{cases}
  1    & r < R_\mathrm{in} \,\, {\rm and} \,\, R_\mathrm{peak} < r \\
  \left(\frac{ r - R_\mathrm{in}}{R_\mathrm{peak} - R_\mathrm{in}}\right)^3 & R_\mathrm{in} < r < R_\mathrm{peak},
    \end{cases}
\end{equation}
where $R_\mathrm{in}$ and $R_\mathrm{peak}$ respectively mark the inner edge and the location of maximum density of the outer ring. We used $R_\mathrm{in,gas}$ = 18\,au and $R_\mathrm{peak,gas}$ = 26.4\,au.

Finally, the total dust-to-gas mass ratio is taken to be $\zeta = 0.047$ \citep[as in][]{Rosenfeld_2013}. The small dust grains are assumed to only make up for a fraction of $f_\mathrm{sd}=1\%$ of the total dust mass. As small dust is typically tightly coupled to the gas dynamics, its density profile is expected to follow the gas density. Then the density of small dust can be calculated as:
\begin{equation}
\rho_{\mathrm{small dust}}(r,z)=\rho_{\mathrm{gas}}(r,z)\, f_{\mathrm{sd}} \: \zeta .
\end{equation}

\subsection{Radial structure: large dust grains}

Since the large dust grains are less coupled to the gas, their distribution has some important differences that require a special parameterisation, such as a larger inner cavity, a larger gap between Ring13 and Ring24, and a break in the outer ring. We only included a low density of large grains within Ring5, just underneath the detection limit of the ALMA observation, as it does not show any visible signature. The surface density profile of the large dust grains is then defined by the sum of its three components
\begin{equation}
  \Sigma_{\mathrm{ld}}(r) = \Sigma_{\mathrm{R5,ld}} + 
  \Sigma_{\mathrm{R13,ld}} + \Sigma_{\mathrm{R24,ld}}.
\end{equation}

For Ring5 and Ring13, we chose Gaussian profiles parameterized with centroid radii $\mu_{\mathrm{R5, ld}}=5.2$\,au and $\mu_{\mathrm{R13, ld}}=13.22$\,au, ring widths of $\sigma_{\mathrm{R5,ld}}=0.1$\,au and $\sigma_{\mathrm{R13,ld}}=0.85$\,au, and normalizations $\Sigma_{\circ,\mathrm{R5,ld}}=1.3\times10^{-4}\,\mathrm{g\,cm^{-3}}$ and $\Sigma_{\circ,\mathrm{R13,ld}}=2.3\,\mathrm{g\,cm^{-3}}$. For Ring24 we used a similar profile as for the gas (a power-law function). The surface density for large dust grains in the outer ring is thus given by
\begin{equation}
    \Sigma_{\mathrm{R24,ld}}(r) = \Sigma_{\mathrm{c}} \left(\frac{r}{R_{\mathrm{c}}(r)}\right)^{-\gamma_{\mathrm{ld}}} \exp\left[-\left(\frac{r}{R_{\mathrm{c}}(r)}\right)^{2-\gamma_{\mathrm{ld}}}\right]\,\delta_{\mathrm{ld}}(r) \,\epsilon(r),
\end{equation}
where for the smoothing factor $\epsilon(r)$ we used $R_\mathrm{in,ld}$ = 24.2\,au and $R_\mathrm{peak,ld} = R_\mathrm{peak,gas}$, resulting in an inner wall of Ring24 at larger radii for the large dust but a peak at the same location than that of the small dust. In an effort to recreate the break seen in the outer ring, we used
\begin{equation}
  \delta_{\mathrm{ld}}(r) =
  \begin{cases}
  0                 & r < 24.6\text{\,au} \\
  1.8 \times10^{5} & 24.6 < r < 27.9\text{\,au} \\
  8.4 \times10^{4} & 27.9 < r < 35.3\text{\,au} \\
  7.1 \times10^{5} & 35.3 < r < 64\text{\,au,} \\
  \end{cases}
\end{equation}
and
\begin{equation}
  \gamma_{\mathrm{ld}}(r) =
  \begin{cases}
  -3.8 & 27.9\text{\,au} < r < 35.3\text{\,au} \\
    1                 &  {\rm elsewhere.} \\
  \end{cases}
\end{equation}

Then the final density of large dust can be calculated as
\begin{equation}
\rho_{\mathrm{large\,dust}}(r,z)=
\frac{\Sigma_{\mathrm{ld}}(r)}{\sqrt{2\pi} \, H(r)} \mathrm{exp}\left[-\frac{1}{2} \left(\frac{z}{H(r)}\right)^2\right] \, f_{\mathrm{ld}} \: \zeta.
\end{equation}
Following the observation, this profile is truncated at 63\,au.

\subsubsection{Reproducing the central emission}
For whole purpose of reproducing the central emission in the ALMA image, we introduce a third and special dust population of larger grains with a distribution tightly confined to the stellar vicinity. This dust is distributed only on a very close-in Gaussian ring parameterized by $\mu_{\mathrm{central\,blob}}=1.1$\,au, $\sigma_{\mathrm{central\,blob}}=0.4$\,au and $\Sigma_{\circ,\mathrm{central\,blob}}=58.8\,\mathrm{g\,cm^{-3}}$, and it is composed of grains with radii ranging from 0.8 to 10\,mm, same upper limit as the large dust but is depleted of small grains. With this distinctive size range we avoid creating a NIR excess in the SED, and we can be consistent with previous mass estimates (see Appendix~\ref{sec:A1} and Sec.~\ref{sec:results}). 

\subsection{Vertical structure}

The parametric scale height profiles for the gas and for each dust population are 
\begin{equation}
    \label{scale}
  H(r)=\chi \, H_{\circ} \,\left( \frac{r}{r_{\circ}}\right)^{1+\psi(r)},
\end{equation}
where $H_\circ$ is the scale height at $r$ = $r_\circ$, $\psi$ is the flaring index and $\chi$ is a scaling factor (in the range $0-1$) that mimics dust settling. In hydrostatical equilibrium, dust diffusion and settling are expected to balance each other, leading to a settling factor of $\chi = \sqrt{D_{\rm d}/(D_{\rm d} + St)}$ \citep{Dubrulle1995}. Here, $D_{\rm d}$ is a dimensionless parameter that informs about the level of diffusivity \citep[it is typically assumed to be similar to the level of turbulence for the particle sizes regarded here][]{2007Icar..192..588Y}. The Stokes number,
\begin{equation}
    {\rm St} = \frac{1}{2} \,\frac{\pi a \rho_{\rm mat}}{\Sigma_{\rm gas}}, 
\end{equation}
summarises the dynamical behaviour of a particle in a given environment (where $\rho_{\rm mat}$ represents a dust particle's material density). By definition, the gas has no settling, and the small dust grains settling is negligible, so that $\chi_{\rm gas}=\chi_{\rm sd}=1$. As we do not know the level of diffusivity in the disc (this will be a topic of interest in Section~\ref{sec:results}), nor the exact value of $\Sigma_{\rm gas}$, we infer the scaling factor for large dust, $\chi_{\rm ld}$, from the width of Ring13. In the radial profile this ring is observed two to three times wider in the gas-tracing NIR than in the fluxes received from larger grains by ALMA. We assume that the same ratio holds in the vertical direction (due to the settling of larger grains towards the mid-plane of the disc), leading to $\chi_{\mathrm{ld}}=0.4$. This is analogous to assuming equal radial and vertical turbulent diffusions.

\begin{figure}
	\includegraphics[width=\columnwidth]{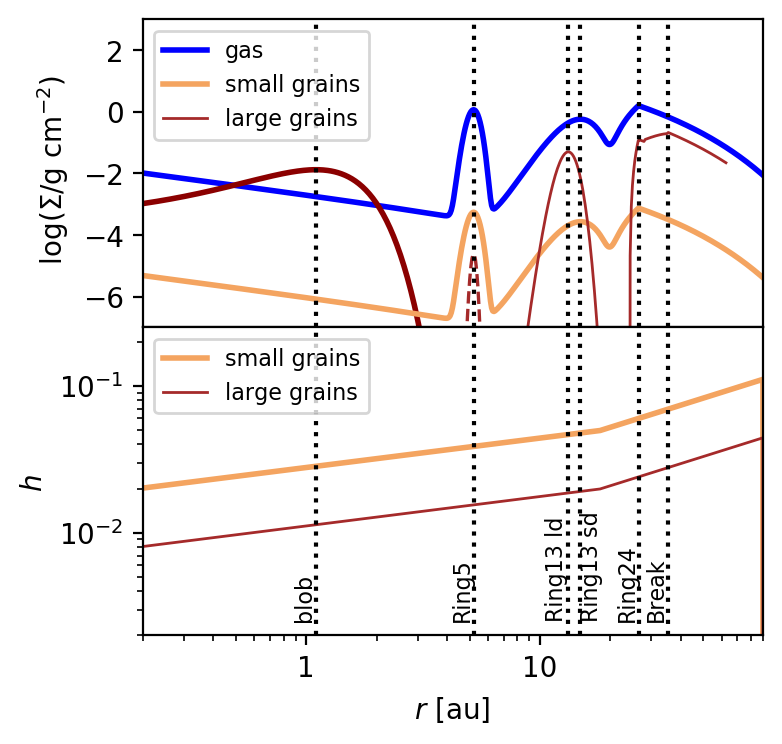}
        \caption{{\bf Top:} surface density profiles for gas, large and small dust grains. {\bf Bottom:} Aspect ratio profile $h(r)= H(r)/r$. The dashed brown line depicts an upper limit for large dust grains within Ring5, inferred from its non-detection in ALMA data. The dotted lines crossing both panels correspond to transition radii in the parametric model.}
    \label{fig:profiles}
\end{figure}

For the vertical structure, \citet{dOrazi} found flaring angles (i.e. height of the ring over the disc midplane divided by the radius of the ring), of $\varphi$ = 6.2$\pm$0.6\,deg for the inner ring and $\varphi$ = 8.5$\pm$1.0\,deg for the outer one. Our model aims to reproduce those values by using two different flaring indices, $\psi_1$ and $\psi_2$. The separation between the two values was set at $r = 18$\,au with $\psi_1=0.2$ for Ring5 and Ring13, and $\psi_2=0.5$ for Ring24. The scale height is set to $H_\circ = 0.89$\,au at $r_\circ = 18$\,au. This vertical structure is consistent with the measurements made by \citet{dOrazi}, as the model flaring angles are $\varphi_{\mathrm{inner}}$ = 6.4\,deg and $\varphi_{\mathrm{outer}}$ = 7.6\,deg. This is summarized in the bottom panel of Fig.~\ref{fig:profiles} that shows the aspect ratio profile $h(r)=H(r)/r$.

\subsection{Image synthesis and SED computation}

To take the simulated images we computed the dust opacities using the {\tt bhmie} code provided in the \textsc{radmc3d} package. The two main populations are taken to be composed by a mix of 60\% silicate, 20\% graphite and 20\% ice. Differently, for the special population of larger grains, as the ring is extremely proximate to the binary system, the dust will not have any traces of ice, so we used a composition mix of 70\% silicate and 30\% graphite.

For the reproduction of the ALMA observation, we create an image using ray-tracing preceded by a Monte Carlo run that gives the simulated image at 1250\,$\micron$, the rest wavelength of the real observation. This image is then convolved with Gaussian blur using the uvmem beam of 0$\farcs021 \, \times \, $0$\farcs018$ to syntesize the final image displayed in Fig.~\ref{fig:images_vs_simulated}. The SED is computed in a similar manner by taking the spectra at 200 wavelengths between 0.1 and 2000\,$\micron$. 

In order to reproduce the \textit{H}-band image, we take a different approach. As the observed asymmetry between the near side and the far side of the disc in the DPI image is suggestive of a strongly forward-scattering phase function, we used much larger grains than typically used in the RT modelling of such NIR data \citep[e.g.,][]{2018MNRAS.477.5104C}. For the computation of this particular image, we implemented a different grain size distribution, where we centred a Gaussian at $\mathrm{a}$ = 0.4\,$\micron$ with $\sigma_{\mathrm{a}}$ = 0.12\,$\micron$ (smeared out by 30\%), and distributed the dust over 20 bins within the range of $\pm \sigma$. This distribution applies only to generate the NIR $Q_\phi$ image and not the ALMA image or the SED. To produce this $Q_\phi$ image we performed a linear combination of the two orthogonal linear polarizations \textit{U} and \textit{Q}, following \citet{Avenhaus_2017}, which gives a representation of an unbiased estimate of the polarized intensity image. The simulated DPI image at 1.65\,$\micron$ in Fig.~\ref{fig:images_vs_simulated} was obtained with the scattering matrix calculated by the {\tt makeopac.py}, script provided in the \textsc{radmc3d} package.

Finally, as mentioned before, the model presented here gives a solution to a highly degenerative problem, and the RT calculation for each set of parameters is very intensive in computational power and time, so a MCMC optimization or similar methods of parameter exploration are impossible to carry out. Nevertheless, as a way to improve the fit and obtain a rough measure of the accuracy of our model, we made one dimensional explorations of the parameter space and found uncertainties of some relevant parameters that will be useful for the discussion (see Appendix~\ref{sec:A1}). We can estimate uncertainties for the scale height at $r_\circ$= 18\,au with $H_\circ = 0.89\pm0.01$\,au and for the width of the gas in Ring13 with $w_\mathrm{g} = 5.30\pm0.27$\,au.

\section{Model results and discussion} \label{sec:results}

The observed SED reveals a small near-infrared (NIR) excess of 0.9$\pm$3.7\,\% \citep{Francis_2020}, this low emission would be primarily emanating from micron-sized dust grains at the hot inner dust wall of a low-mass inner disc. On the other hand, the mm-bright central emission suggests a massive inner dust ring. The faint near-IR excess contrast with the bright mm emission, and could point to a lack of micron-sized dust in the central ring, that could be due to efficient dust growth or due to the ring having an inner radius well beyond the sublimation radius. In order to reproduce the low NIR and simultaneous bright central mm emission, we extended the radius of an inner cavity to significantly exceed both the sublimation radius and the zone that is expected to be cleared by dynamical binary-disc interaction (see Appendix~\ref{sec:A2}). A possible explanation for this wider cavity is the presence of an additional effect of truncation, like an unseen companion planet in the very inner part of the disc \citep{Francis_2020}. At the same time, the central emission is also well reproduced in the ALMA image with the inclusion of a Gaussian ring at 1.1\,au. This inner ring produces a low near-IR excess only if it consists of dust grains larger than $\sim$0.3\,mm (see Appendix.~\ref{sec:A1}). But we implement 0.8\,mm instead as the lower limit for the dust population that composes this feature, given that this predicts a dust mass of 0.012\,M$_{\earth}$, which is close to that obtained by \citet{Francis_2020} (0.013$\pm$0.002,M${\earth}$). They converted mm-flux into mass using the standard opacity value of $\kappa_{\nu}$=10\,$\mathrm{cm}^2 \mathrm{g}^{-1}$ at 1000\,GHz with an opacity power-law index of $\beta$=1.0, while our mass estimates are  extracted from the RT model.
\begin{figure}
	\includegraphics[width=\columnwidth]{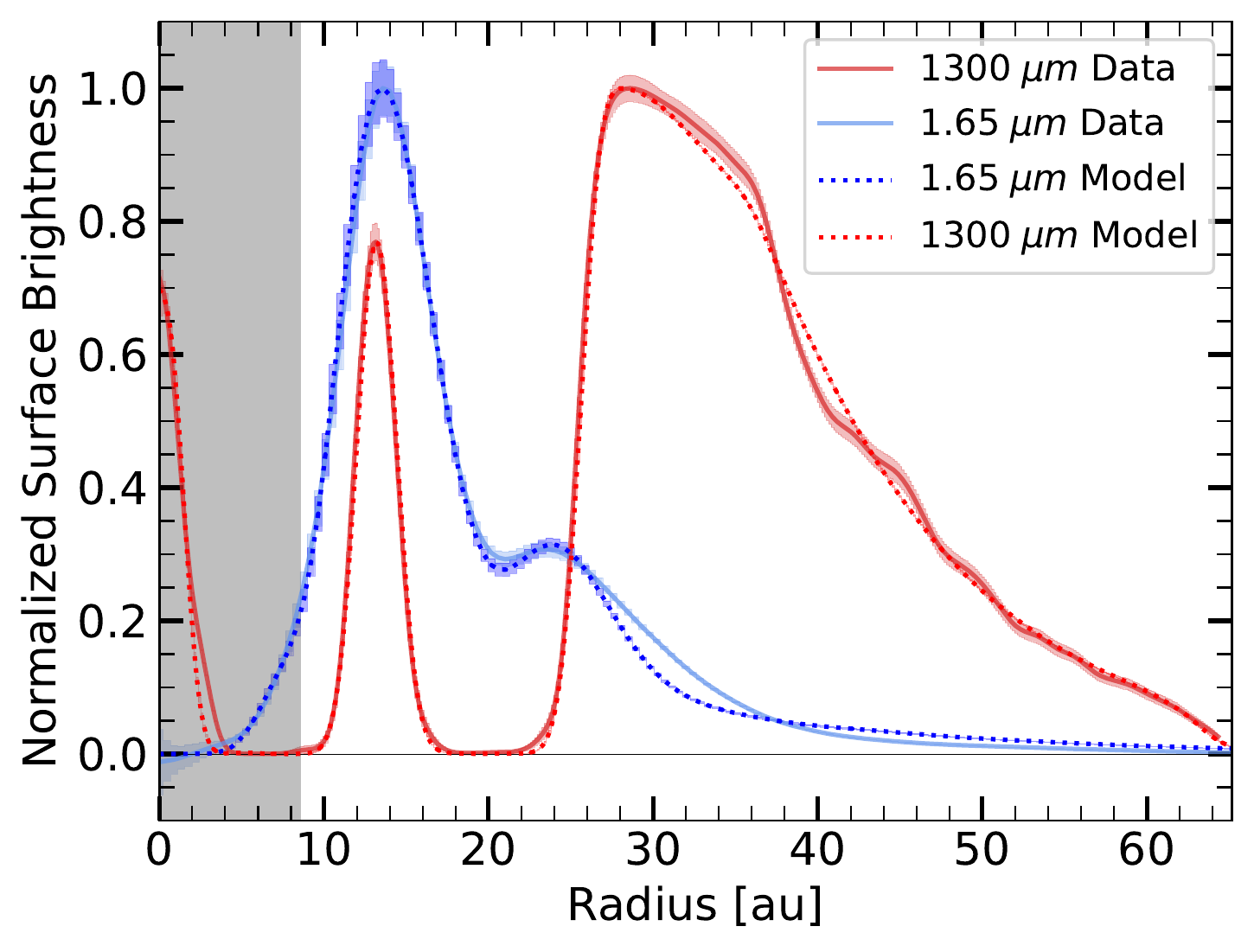}
    \caption{Comparison of the surface brightness profiles extracted from the deprojected synthetic images and observed \textit{H}-band and 1.3\,mm continuum images. The grey shaded area represents the radius of the artificial coronagraph used in the simulations (i.e., $\sim$0$\farcs 1$, or $\sim$7.1\,au at 71.48\,pc).}
    \label{fig:radprofiles}
\end{figure}

The decision of including Ring5 to the observed structure relies on the fact that the SED needs a thin dust ring, made mainly of small dust grains, at a radius of $\sim$5\,au to have a proper fit between 6 and 300\,$\micron$ (see Appendix~\ref{sec:A2}). The introduced Ring5 is not visible in the simulated image at 1.65\,$\micron$, where it hides under the artificial coronagraph, neither in the 1.3\,mm continuum simulated image, as its predicted peak intensity is around two times the noise in the ALMA image ($\sim$1$\times10^{-7}$\,Jy\,$\mathrm{beam}^{-1}$). This gives us an upper limit for the total millimetre--sized dust mass present in Ring5 of $\sim$2$\times10^{-5}$\,M$_{\earth}$. The depletion of large dust grains in Ring5 is consistent with efficient dust trapping in Ring13. A zone of radially increasing gas pressure can entirely filter out large dust grains from the inner regions, while smaller grains can be able to overcome this barrier due to their strong frictional coupling to the gas accretion flow \citep[studied in the context of planetary gaps,][]{Rice2006,Zhu2012,Weber2018}. Still, the strong depletion of large particles demands that within Ring5 dust growth is extremely inefficient or limited by fragmentation. Otherwise, the present small grains should coagulate to form detectable grain sizes in the ALMA observation in Fig.~\ref{fig:images_vs_simulated} \citep{Drazkowska2019}. 

As the radial profiles obtained from the simulated images of the model closely resemble those deduced from the observations (Fig.~\ref{fig:radprofiles}), we can assume that the model provides a possible approximation of the disc structure, including the dimensions of Ring13. The FWHM of the gas and micron-sized dust in Ring13 in the RT model corresponds to $w_\mathrm{g} = 5.30\pm0.27$\,au, as well as a radius of 14.9\,au, and a scale height FWHM of 0.63\,au. Meanwhile, the millimetre-sized dust in Ring13 has a FWHM of 2.00\,au, a radius of 13.22\,au, and a scale height FWHM of 0.25\,au ($\sim2.355\times H(\mu_\mathrm{R13,ld})$). The total dust mass of Ring13 would be about 0.7\,M$_{\earth}$. The model predictions for the millimetre-sized dust population in Ring13 are close to the measurements, with only a 1\% difference in the centroid location of the ring, and a 19\% difference in the width estimation. Given that the scale height FWHM of the large grains in Ring13 is 0.25\,au, and that the width of Ring13 from the ALMA observations is 2.46$\pm$0.56\,au (see Sec.~\ref{subsec:ALMA}), we conclude that the large dust ring is 10.0$\pm$1.6 times more extended radially than vertically. Looking at the rest of the disc, our model reproduces the observations of Ring24 with peak intensity at $\sim$30\,au, and the break at $\sim$36\,au. The whole disc contains a total dust mass of $\sim$48\,M$_{\earth}$.

Even though the radial spread of large dust grains in Ring13 appears to be quite thin, the width in comparison to the sub-lying gas profile speaks for the presence of considerable turbulent diffusion. Following a similar ring analysis as in \citet{2018ApJ...869L..46D}, we find that the ratio between the dimensionless diffusion parameter, $D_{\rm d}$, and the dimensionless Stokes number, St, (which parameterizes the dynamical behaviour of a grain) is roughly, $D_{\rm d}/{\rm St}\approx 0.1$. The observed signal is expected to be dominated by grains of size $a\approx 0.02\,{\rm cm}$. The RT model, together with the dust-to-gas ratio of 0.05, prescribe a gas density of $\Sigma_{\rm g} \approx 0.5\,{\rm g}{\rm cm}^{-2}$ to the location of Ring13. With these values, the relevant Stokes number is approximated to be St$\approx 0.1$. This yields an estimate for the level of diffusivity of $D_{\rm d}\approx 0.01$. It further provides a value for the level of turbulent viscosity in Ring13, $\alpha_{\rm turb}\approx 0.01$, assuming the level of turbulence to be equal to the level of diffusion \citep{2007Icar..192..588Y}. We note that an observation of molecular line broadening has found no evidence for turbulent contributions, suggesting $\alpha_{\rm turb}<0.01$ \citep{Flaherty_2020}. The value inferred from our model is just within this limit. By our definition of the gas surface density profile, the value inferred for the level of turbulence is linearly proportional to the local dust-to-gas ratio. Lower values than the chosen ratio of 0.05 would, therefore, lead to an equally lower level of turbulence in the assessment. While the exact value for $\alpha_{\rm turb}$ is not well constrained, a certain level of turbulence is required to explain the radial spread of the resolved Ring13.

The visible asymmetry in the SPHERE observations is reproduced using relatively large grains, $\sim$0.4\,$\micron$, as smaller grains did not result in such strong forward scattering. As \citet{refId0} state, the strong forward scattering that is present in the observation may indicate that the dust grains in the disc surface are relatively large, suggesting that the disc is depleted of very small grains. Alternatively, it may suggest that grains are not spherical as assumed in the calculations of opacities using Mie theory. 

Another interesting feature of the simulated 1.65\,$\micron$ image is that the model accurately shows the shadows described by \citet{dOrazi} that are present in the SPHERE–IRDIS image. In contrast, there are no hints of radio decrements along Ring13 or in Ring24, in either the ALMA observations or in the simulated 1.3\,mm continuum image, that would match the shadows. As noted by \citet{Casassus2019MNRAS.486L..58C}, the diffusion of thermal radiation from the disc smooths out the decrements seen in scattered light, and in this case it is likely that the disc cooling time-scale is much slower than the that of the illumination pattern.

The general observed structure may point to the existence of planet-disc interactions within this system, where giant planets deplete their orbits of gas and dust material. A possible planetary constellation in this scenario is, therefore, the presence of two giant planets in the disc, one planet between the star and Ring13, and one planet between Ring13 and Ring24. As \citet{Ru_z_Rodr_guez_2019} suggest, the putative planet between Ring13 and Ring24 may be a giant planet with a mass within the range of 0.3$-$1.5\,$\mathrm{M}_{\mathrm{Jup}}$. This idea is supported by a dedicated study (Weber et al. submitted) which qualitatively reproduces the observations of this system with a hydrodynamical simulation including several giant planets.

The expected age of >20\,Myr of V4046\,Sgr suggests that its gas-rich disc is unconventionally old in comparison to typical circumstellar examples \citep[e.g.][]{Fedele2010, Williams2011}. The dispersal of such gas discs is typically assumed to be set by photo-evaporation \citep{Alexander2006,Gorti2009}. While the dynamical origin of the disc’s longevity is not the subject of the present study, we would like to mention that its occurrence around such a close binary might not be coincidental. \citet{Alexander2012} predicted that disc lifetimes should show a sharp increase around binaries separated by $\lesssim 0.3-1.0\,$au. It still has to be seen whether a trend towards longer disc lifetimes in compact multiple-star systems (as recently proposed by \citealp{Ronco2021}) turns out to be prevalent.

\section{Conclusions} \label{sec:Conclusions}

We present new ALMA 1.3\,mm continuum imaging of V4046\,Sgr, a well-known circumbinary disc, at an unprecedented definition ($\sim$0$\farcs021 \, \times \, \sim$0$\farcs018$), where new features become visible. Together with the analysis of a SPHERE--IRDIS polarized image and a well-sampled SED, we aim to reproduce the observations with radiative transfer modeling, looking for a way to explain the data in terms of a physically model. The key conclusions of this analysis are as follows.
\begin{enumerate}
    \item The central emission in the ALMA image suggests the existence of an inner ring of dust grains larger than 0.8\,mm. Our interpretation agrees with the mass estimation of this feature made by \citet{Francis_2020}, with a mass of 0.012\,M$_{\earth}$.
  
    \item Our parametric model, which accounts for the SED of the system, predicts the presence of an inner ring at $\sim$5\,au, mainly consisting of small dust grains. This additional ring lies under the coronagraph of the scattered light image and is too faint to be detected by the ALMA observation. The depletion of large dust in this ring is consistent with efficient dust trapping at larger radii, as can be expected in Ring13. 
  
    \item The narrow ring in the 1.3\,mm continuum, has a radius of 13.15$\pm$0.42\,au and an estimated width of 2.46$\pm$0.56\,au. The location of this ring is coincident with the inner ring observed in the scattered light image. From our RT modeling we can predict that this ring includes around 0.7\,M$_{\earth}$ of millimetre-sized grains. Using the parametric model scale height FWHM value for the large grains ($H_{\mathrm{ld}}=$ 0.25\,au at 13.15\,au) we find that the ring width is roughly 10 times its estimated height. 
    
    \item The 1.3\,mm outer ring, that starts at $\sim$24\,au and has its peak intensity at $\sim$30\,au, presents a visible break in the surface brightness at $\sim$36\,au.
  
    \item While we can not get an exact value for $\alpha_{\rm turb}$, the resolved radial width of Ring13 speaks for the presence of a considerable level of turbulent viscosity.

    \item We interpret the asymmetry observed with SPHERE--IRDIS at 1.65\,$\micron$ as due to strong forward-scattering, which implies that the dust population is depleted of grains smaller than $\sim$0.4\,$\micron$.
  
\end{enumerate}

\section*{Acknowledgements}

We thank the anonymous referee for their constructive review. R.MB., S.C., S.P., L.C., and A.Z. acknowledge support from FONDECYT grants 1211496, 1191934, 1211656 and 11190837.
P.W. acknowledges support from ALMA-ANID postdoctoral fellowship 31180050. M.C. acknowledges support from ANID PFCHA/DOCTORADO BECAS CHILE/2018-72190574. 
This paper makes use of the following ALMA data: ADS/JAO.ALMA \#2017.0.01167.S. ALMA is a partnership of ESO (representing its member states), NSF (USA) and NINS (Japan), together with NRC (Canada), MOST and ASIAA (Taiwan), and KASI (Republic of Korea), in cooperation with the Republic of Chile. The Joint ALMA Observatory is operated by ESO, AUI/NRAO and NAOJ. The National Radio Astronomy Observatory is a facility of the National Science Foundation operated under cooperative agreement by Associated Universities, Inc. This research has made use of the VizieR catalogue access tool, CDS, Strasbourg, France (DOI : 10.26093/cds/vizier). The original description of the VizieR service was published in \citet{2000A&AS..143...23O}. This research has made use of the NASA/IPAC Infrared Science Archive, which is funded by the National Aeronautics and Space Administration and operated by the California Institute of Technology.

\section*{Data Availability}

The data used in this article are presented in Fig.~\ref{fig:images_vs_simulated}. The final images can be found in FITS format in supplementary materials while the original raw data can be downloaded directly from the ALMA archive using project code 2017.1.01167.S. The SPHERE/IRDIS data can be downloaded from the ESO archive using project code 096.C-0523(A).



\bibliographystyle{mnras}
\bibliography{bibtex} 



\appendix

\section{Parameter--space exploitation} \label{sec:Appendix}

\subsection{Parameter space partial exploration} \label{sec:A1}

\begin{figure*}
	\includegraphics[width=\textwidth]{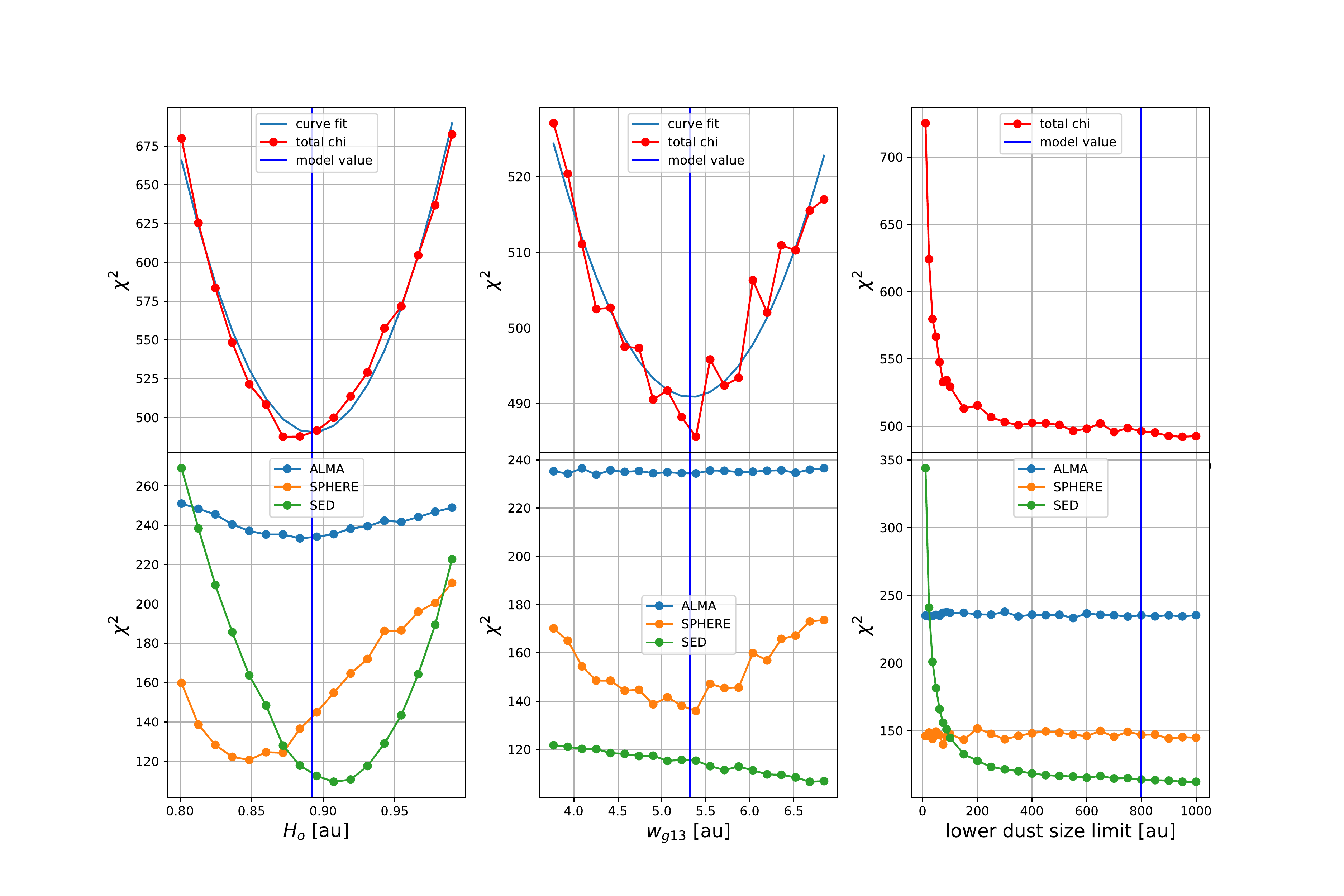}
        \caption{One dimensional explorations of the chi squared space of our parametric model. The first panel shows the variation of $\chi^2$ in terms of the scale height $H_\circ$ at $r_\circ$=18\,au, the second one for the width of the gas component of Ring13 ($w_{\mathrm{g13}}$), and the third one around the lower limit of the grain size in the close inner ring. Each panel displays at the bottom the $\chi^2$ values for the SED and for the two images separately, and on top the total $\chi^2$ value in terms of the parameter value.}
    \label{fig:chi}
\end{figure*}

As a way to compute a measure of goodness of fit and to try to quantify the model uncertainties, we performed a partial exploration of the parameter space. 

We explore $\chi^2$ space in the vicinity of the model values (obtained by trial and error, see Sec.~\ref{sec:model}) for the scale height at $r_\circ$ ($H_\circ$), the width of the gas component of Ring13 $w_{\mathrm{g13}}$, and around the lower limit of the grain sizes in the central blob. The parameter space exploration is shown in Fig.~\ref{fig:chi}. Our model fits the SED and two images, so the total $\chi^2$ value for a given model is composed of the sum of the $\chi^2$ values for each of the three fits.
\begin{equation}
    \chi^2_{\mathrm{total}}=\chi^2_{\mathrm{SED}}+\chi^2_{\mathrm{ALMA}}+\chi^2_{\mathrm{SPHERE}}.
\end{equation}

For $H_\circ$ and $w_{\mathrm{g13}}$, we find that the values of the parameters in our model are at a minimum in each one dimensional $\chi^2$ space, the uncertainties (up and down) will be those that correspond to $\chi^2=\chi^2_\mathrm{m}+1$, where $\chi^2_\mathrm{m}$ is the local minimum value. We approximated the vicinity of the local minimum in these 1D cuts with a quadratic fit.

For the lower grain size limit, we found that despite that our chosen value seems to minimise $\chi^2$, the shape that follows the curve is more suggestive of a border condition. So there is a threshold around 300\,$\micron$ where the SED starts to deviate strongly from the typical value. That point represents where the NIR excess becomes significantly larger that the observed.

\subsection{Comparison between different models}\label{sec:A2}

\begin{figure}
	\includegraphics[width=\columnwidth]{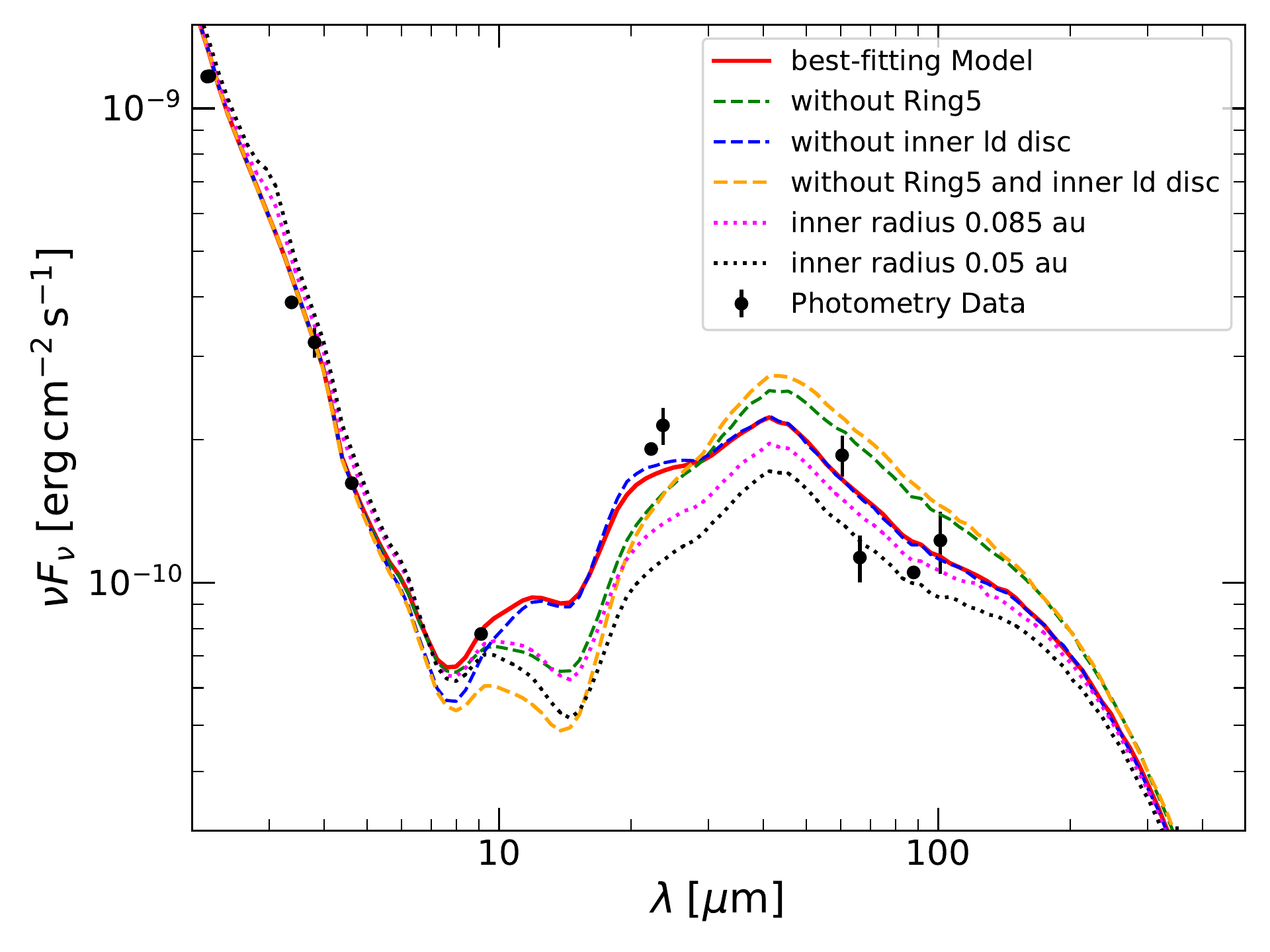}
        \caption{{Comparison between the SEDs of best-fitting model versus five other models with differences in the dust structure. The dotted lines represent the models with different inner radius while the dashed lines represent the models without Ring5 or the inner large-dust disc.}}
    \label{fig:SED_comparition}

\end{figure}

A compact inner hole in the dust distribution around a binary can typically be produced by sublimation of solids or dynamical clearing of the central stars. The sublimation radius of the system is expected to be at $\sim$0.05\,au \citep[$R_{\mathrm{sub}}=0.07\sqrt{L_*(L_{\sun})}$\,au,][]{Francis_2020}, and the edge of the zone cleared by dynamical interactions between the near-circular binary and the circumbinary disc is estimated to be at 0.085\,au \citep[$r=2.08a$,][]{Art_Lu}. On the other hand, for V4046 Sgr, \citet{Jensen_97} required to implement a cavity to 0.2\,au to fit the SED flux around the silicate feature at 10\,$\micron$. We find that the latter value significantly improves the SED fit when compared to a disc extending to the inner radius predicted by dynamical clearing in the SED fit (see Fig.~\ref{fig:SED_comparition}), therefore, the inner radius of the disc in our best-fitting model lies at 0.2\,au.

The final structure of our proposed disc also includes additional features that are necessary to fit both the SED and the images. These features are a cut-off of the inner density accumulation of large dust grains outside of 1.1\,au (to fit the central emission in the ALMA image), and the thin Gaussian ring made mainly of small dust grains at 5.2\,au (Ring5, demanded by mid-infrared excess in the SED). The first one is visible in the ALMA image so we are forced to include it, but Ring5 is not detected in the ALMA continuum and was included because of its contribution to the 10$\micron$ flux is required by the SED. Fig.~\ref{fig:SED_comparition} works as a summary as it shows a comparison between our best-fitting model and five models with changes on their structures: one without Ring5, one without the inner large-dust disc, one without both of those structures, another with an inner edge of the disc at 0.085\,au, and a last one with an inner edge of the disc at 0.05\,au. All these models diverge from the SED data somewhere in the 6-300\,$\micron$ range.

\bsp	
\label{lastpage}
\end{document}